\documentclass[showpacs,eqsecnum,twocolumn]{revtex4}%preprint,
\usepackage{xspace,psfig,epsfig,amsfonts,amssymb}

\newcommand{\beq}{\begin{equation}}
\newcommand{\eeq}{\end{equation}}
\newcommand{\bqa}{\begin{eqnarray}}
\newcommand{\eqa}{\end{eqnarray}}
\newcommand{\nn}{\nonumber}
\newcommand{\nl}[1]{\nn \\ && {#1}\,}
\newcommand{\erf}[1]{Eq.~(\ref{#1})}
\newcommand{\dg}{^\dagger}
\newcommand{\rt}[1]{\sqrt{#1}\,}
\newcommand{\smallfrac}[2]{\mbox{$\frac{#1}{#2}$}}
\newcommand{\half}{\smallfrac{1}{2}}

\newcommand{\hei}{Heisenberg\xspace}

\newcommand{\ito}{It\^o\xspace}

\newcommand{\sq}[1]{\left[ {#1} \right]}
\newcommand{\cu}[1]{\left\{ {#1} \right\}}

\newcommand{\an}[1]{\left\langle{#1}\right\rangle}

\newcommand{\raro}{\rightarrow}
\newcommand{\col}{0.45\textwidth}
\newcommand{\erfs}[2]{Eqs.~(\ref{#1})--(\ref{#2})}

\newcommand{\gamr}{\gamma_{{\rm r}}}
\newcommand{\dtime}{\tau_{{\rm dd}}}
\newcommand{\dwj}{d{\cal W}_{{\rm J}}(t)}
\newcommand{\dwjp}{dW_{{\rm J}}(t)}

\begin{document}

\title{Quantum Trajectories for Realistic Photodetection I: General
Formalism}

\author{P. Warszawski}
\affiliation{Centre	for	Quantum	Dynamics, School of	Science,
Griffith University,
Brisbane 4111, Australia.}
\author{H. M. Wiseman}
\email{H.Wiseman@gu.edu.au}
\affiliation{Centre	for	Quantum	Dynamics, School of	Science,
Griffith University,
Brisbane 4111, Australia.}

\begin{abstract}
Quantum trajectories describe the stochastic evolution of an open quantum
system conditioned on continuous monitoring of its output, such as by an
ideal photodetector.  In practice an experimenter has access to an output
filtered through various electronic devices, rather than the microscopic
states
of the detector.  This introduces several imperfections into the measurement
process, of which only inefficiency has previously been incorporated
into quantum trajectory theory. However, all electronic devices have finite
bandwidths, and the consequent delay in conveying the output signal
to the observer implies that the evolution of the conditional 
state of the quantum system must
be non-Markovian.  We present a general method of describing this
evolution and apply it to avalanche photodiodes (APDs) and to
photoreceivers. We  include the effects of efficiency, dead
time, bandwidth, electronic noise, and dark counts. The essential idea
is to treat the quantum system and classical detector jointly, and to
average over the latter to obtain the conditional quantum state.
 The significance of our theory is
that quantum trajectories for realistic detection are necessary for
sophisticated approaches to quantum feedback, and our approach could be
applied in many areas of physics.
\end{abstract}
\pacs{03.65.Yz, 03.65.Ta, 42.50.Lc, 42.50.Ar}

\maketitle

\section{Introduction}

\subsection{Measurement in Quantum Mechanics}

To obtain information about a system of interest a measurement has to
be made.  In
experiments that probe the quantum nature of our world, the
system itself is, in general, necessarily affected by the act of measurement.
For experiments involving continuous  monitoring, traditional
(projective) measurement theory as axiomatized by von
Neumann \cite{Von32} is inadequate as a description.
Such measurements would
prevent any interesting evolution occurring, because of the
quantum Zeno effect \cite{zeno1}.  To avoid this, a generalized
quantum theory of measurement must be used \cite{Dav76,Kra83}.
Generalized quantum measurements
can be derived by considering projective measurements on a second
system interacting with the system of interest.
Often the second system has a clear physical interpretation.
For example, in quantum optics a bath  (continuum)
of electromagnetic field modes plays this role and the quantum system
is said to be {\em open}.

If the system is weakly coupled to its bath and the
dynamics are such that information concerning the system is spread
throughout the many degrees of freedom of the bath, then a Markovian
evolution equation for the system alone may be obtained.
 This equation is known as a {\em master equation} (ME).
If, in addition, the bath is being measured then a stochastic
master equation (SME) for the conditioned state of the system can be found.  This is
termed a {\em quantum trajectory} \cite{Car93b}. The relation of the ME
to quantum trajectories is that the former results from summing all
possible trajectories, weighted by their probability of
occurrence.  Thus, the ME represents average
evolution and gives the evolution of the state of the system when the
environment is not monitored. A
quantum trajectory (which is said to `unravel' the master equation)
gives the evolution of the state of the system
conditioned on the results of monitoring the environment.
A quantum trajectory is necessarily stochastic, and 
can be jumpy \cite{Car93b,DalCasMol92,GarParZol92}
or diffusive \cite{Bel88,BelSta92,Bar90,Bar93,Car93b,WisMil93c},
 depending on the nature of the 
measurement.

\subsection{Realistic Quantum Trajectories}

In experiments performed in the laboratory, a detector
mediates between the detector input (for example, the  electromagnetic
field emitted by a fluorescent quantum system) and the experimenter.  
Realistic detectors are not perfect.  Information is lost in the 
conversion of the quantum field 
to a signal that the observer can use.  
This loss may occur at
the front end of the detector (characterized by an inefficiency), in
the circuit containing the detector (described by a response time and
electronic noise) or at the circuit output (electronic output noise).
In order to obtain a true quantum trajectory for the
experiment, the observer must condition the state of the quantum
system on results that are available in the laboratory rather than on the
microscopic absorption
events considered previously in quantum trajectories.

Realistic quantum trajectories of this sort cannot be encompassed
within the standard formalism of
Refs.~\cite{Dav76,Kra83}, except for inefficiency \cite{WisMil93a}.
This is because the electrical circuit,
which filters the measured output of the quantum system, causes a
statistical delay in the observer obtaining information about the system.
Thus the conditioned evolution of the quantum system
alone is non-Markovian.  A method for treating this was first proposed
by us and Mabuchi~\cite{WarWisMab01}, 
in which the quantum system is embedded within a
supersystem that also contains the state of the detector.
If the set of (classical) detector states is $\mathbb{S}$, then the
supersystem
is described by the set $\{\rho_{s}:s \in \mathbb{S}\}$. Here ${\rm
Tr}[\rho_{s}]$ is the probability that the apparatus is in state $s$,
and $\rho_{s}/{\rm Tr}[\rho_{s}]$ is the system state
given this event.  The supersystem state, $\cu{\rho_{s}}$, obeys a
Markovian evolution equation.  The detector states are classical in the sense 
that it is not necessary to 
consider superpositions between them.

In this paper we consider in detail two sorts of realistic detectors,
an avalanche photodiode in geiger mode (a photon counter) and a
photoreceiver (for homodyne detection). The detector states in the
former case are discrete (we use three states) and in the latter 
continuous  (although discretization is necessary for numerical
simulation). For the  photon counter, this model allows us to
include the effects of detection
efficiency, dark counts, response (or rise) time and dead time.  For homodyne
detection using a photoreceiver we investigate detector efficiency,
output Johnson noise (which adds onto the measurement result linearly)
and detector bandwidth. (Of course, this bandwidth
  relates not to the range of frequencies
(colours) of light to which the detector will respond, but to the
characteristic time it takes for the absorption of light by the detector to
generate a response in the detector output.)

Before passing onto our method of derivation of realistic quantum
trajectories, it is  worth mentioning that a three-level model
for a photon counter
including many of the above effects has been considered by Gardiner
\cite{Gar91}. However this theory is based upon quantum Langevin
equations rather than quantum trajectories,
and does not allow one to determine the system state conditioned on
the experimenter's knowledge.

\subsection{Derivation of Realistic Quantum Trajectories}
\label{Deriv}
\begin{figure}
\includegraphics[width=\col]{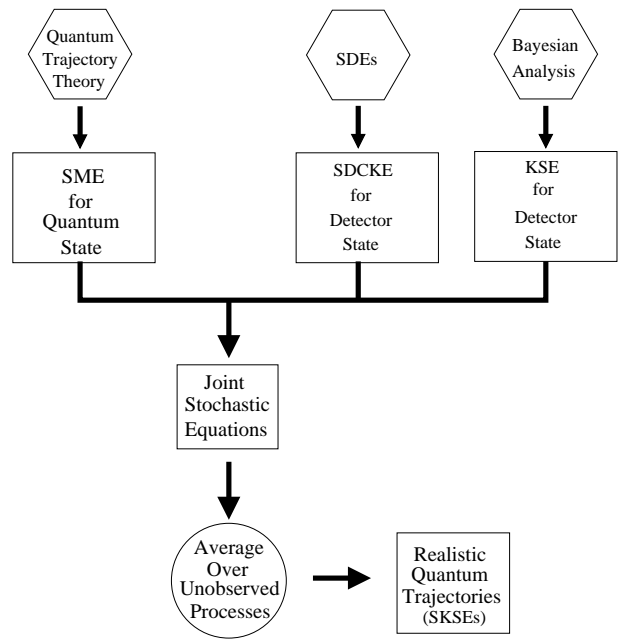}
\vspace{0.2cm}
\caption{Schematic diagram for the method that
we use to derive realistic quantum trajectories.  It should be noted
that some of the steps shown above can be performed in a different
order.  This flexibility is useful for keeping the derivation of
realistic quantum trajectories as simple as possible.  The following
abbreviations have been used: stochastic master equation (SME),
stochastic differential equation (SDE), stochastic differential
Chapman-Kolmogorov equation (SDCKE), Kushner-Stratonovich equation (KSE)
and superoperator Kushner-Stratonovich equation (SKSE).
Hexagons, rectangles and circles indicate bodies
of knowledge, equations and procedures, respectively.
See text for further details.}
\protect\label{Method}
\end{figure}

The derivation of realistic quantum trajectories consists of a number
of well defined steps, that are schematically illustrated in
Fig.~\ref{Method}.  These can be understood by considering the
subsystems that make up the supersystem, namely the quantum
system and the detector.

The basic philosophy
that we adopt at the start of the derivation is that {\em in
principle} all of the microscopic states of the detector could be
monitored, without altering the quantum trajectory of the system.
Essentially this is because the detector states consist of distinct
configurations of an
enormous number of particles. These particles are constantly interacting with
their environment, thus making coherences between detector states unimportant.
Thus in principle it is possible for an experimenter to have
sufficient information to unravel the evolution of
the system by a conventional stochastic master equation (SME).
In practice, an experimenter has access to far less knowledge. This
loss of knowledge is due to {\em classical} uncertainties
introduced by the detector.

To describe the detector we consider some property of it
that can be related to the output that the observer monitors.  For
example, in the case of homodyne detection the charge across one of
the capacitors in the circuit represents the detector state.  By
considering all of the dynamic influences upon this circuit variable we
formulate an evolution equation for it that is stochastic in general.
Since finally
we will not know the stochastic inputs, we transform this
into a stochastic
differential Chapman-Kolmogorov equation (SDCKE) that
follows the probability of the various states being occupied.
We use this terminology to indicate a generalization of the
deterministic differential Chapman-Kolmogorov
equation discussed in
\cite{Gar85}.

The realistic measurement results available to the observer give
information about the detector state, without necessarily revealing
it fully. Based on the result and the {\em a priori} (initial) probability
distribution for the detector
state, the best estimate of the updated state (the {\em a
posteriori} distribution) is found.  This is done via a Bayesian analysis
\cite{BayesBook}.  The resulting equation for the the probability
distribution is, for the case of Gaussian white noise, usually
 referred to as a Kushner-Stratonovich
equation (KSE) \cite{KSBook}. 
For convenience we will use this terminology even in
the case of point-process noise.

Note that normally a KSE would also contain that
unconditioned (deterministic) evolution due to the system dynamics.
Since these dynamics are stochastic at this stage of the derivation
(they are contained in the SDCKE),
we do not include them in the KSE for the detector state. 
Note also that the SME, despite its name (introduced
in Ref.~\cite{WisMil93a}), is conceptually closer to the KSE than to
the SDCKE: the stochasticity in both the KSE and the SME is solely due to
measurement noise, and the stochastic terms are nonlinear in both
cases. In fact, the SME can justifiably be called a quantum KSE
\cite{Dohe2000}.

The evolution of the SME, the SDCKE and the KSE
are combined to give the joint stochastic equations governing the
evolution of the supersystem.  Up until this stage all the stochastic
influences on the supersystem are still present.  To link the
evolution of the system with that of the detector this is necessary.  Now,
however, we average over processes that a realistic observer cannot
monitor.  This leaves only the stochasticity associated with the
measurement and our derivation of the realistic quantum trajectory is
then complete.  We term this final stochastic trajectory for the supersystem
a superoperator Kushner-Stratonovich equation (SKSE).  
Obviously the reader will gain much greater insight
into this process by following the examples provided in the later
sections.

\subsection{Importance of Realistic Quantum Trajectories}

Thus far, the main practical utility of quantum trajectory theory has been in
improving the computational efficiency of simulations used to compare models
with experimental data. For this purpose, the introduction of
realistic quantum trajectories could only hinder simulations. However,
quantum trajectory theory is now gaining increasing importance as the
quantum generalization of Kalman filtering, which provides essential
signal-processing methods in classical estimation, communication, and control
engineering. Quantum trajectory theory should in principle play the same
pivotal role for emerging quantum analogs of these technologies
\cite{Gamb01,Cira97,Dohe99b}. Before this can happen it is essential that the
theory be extended to account for the imperfections of realistic
measurement devices. Non-ideal detector dynamics can dramatically affect
the proper inference from measured signals to the conditional quantum state
of an observed system, as we will show.

\subsection{Overview of this Paper}

As stated above, the main thrust of this paper is to derive in detail
the realistic quantum trajectories for photon counting and homodyne
photodetection introduced in Ref.~\cite{WarWisMab01}.
In Sec.~\ref{Theory}, we first briefly give the theory of classical
trajectories and measurement, which is important for obtaining SDCKEs
and KSEs for the detector state.  A discussion of
non-realistic (that is, ideal)
quantum trajectories follows this, which allows the SMEs for the
quantum system to be found.

In Secs.~\ref{PCounter} and \ref{Photoreceiver}, 
realistic quantum trajectories are
derived for photon counters and photoreceivers respectively.  Both of these
derivations are based on the procedure outlined in Sec.~\ref{Deriv}.
For the photon counter, we finally idealize the measurement process to
show that our equations reduce to those associated with perfect
detection (traditional quantum trajectories).
An important and non-trivial consideration in the case of the photoreceiver
is the bandwidth of the device.  As the photoreceiver circuit adds
electronic noise, a calculation of this bandwidth based on just the
resistance-capacitance ($RC$) time constant of the circuit does not yield the
correct answer.
As a final point we consider the effect of letting the $RC$ time
constant go to zero but still maintaining the presence of the
electronic noise which obscures the input signal.
We conclude with a summary of our results and discussion of future
directions in Sec.~\ref{Concl}. 
A numerical exploration of our theory is given in the following paper 
\cite{WarWis02b}.

\section{Stochastic Equations}
\label{Theory}
As discussed in Sec.~\ref{Deriv}, there are three types of stochastic
equations that go into making a realistic quantum trajectory. In this
section we examine the general theory for each.

\subsection{Stochastic Differential Chapman-Kolmogorov Equation}
\label{Trajrefer}

A typical Langevin equation or stochastic differential equation (SDE)
for a classical random variable $X$ can be written as
\beq
dX=a(X)dt+\rt{D}dW(t)+edN(t).
\label{Langevin}
\eeq
Here $a(X)$ is an arbitrary function of $X$,
and $dW(t)$ and $dN(t)$ are  stochastic increments.
The point process $dN(t)$   is
either zero or one, and has an infinitesimal mean which is a positive
function of
$X$, say $g(X)dt$. The Wiener increment  $dW(t)$  is
related to Gaussian white noise $\xi(t)$ by $dW(t)=\xi(t)dt$.
Gaussian white noise has statistics of \cite{Gar85}
\beq
{\rm E}[\xi(t)]=0\label{xiMean},\;\;
{\rm E}[\xi(t)\xi(t')]=\delta(t-t').\label{xiSquare}
\eeq
Here the roman font E denotes expectation value.
By writing the SDE in the form of \erf{Langevin} we imply that the
equation is to be interpreted in the \ito
\cite{Gar85} or explicit \cite{Wis94} sense. This means, for example,
that the expected value of $dX$ may be found by replacing $dN$ and
$dW$ by their
expected value. This is not true for the implicit \cite{Wis94}
or Stratonovich \cite{Gar85} form. By the
same token, the  chain rule of standard
calculus does not apply in general to the \ito form, whereas it does for the
Stratonovich form.

Langevin equations  are
appropriate for situations in which the stochastic increment is
known, thus allowing $X$ to be tracked.  Often it is relatively easy
to turn the physics of a problem into a Langevin equation.  An
example of this is a noisy electrical circuit.  The analysis of such
circuits is well understood and is based on simple principles such as
Kirchhoff's laws.

In this paper, Langevin equations are first found on
the {\em presumption} that the stochastic increment is known.  In
reality, however, this is not the case and we cannot track the
variable $X$ perfectly.  For this reason it is necessary to
turn the Langevin equation into an equation describing the evolution
of the {\em probability distribution} for $X$, which is written as
$P(x)$.  Note that the
transformations that we perform actually maintain the stochastic nature of
this evolution equation, as the averaging over unobserved processes is
done at a later stage. This is necessary because the noise in the
equation for $P(x)$ may also appear in other parts of the overall derivation.
Before this final averaging is done, the stochastic equation for $P(x)$,
 would preserve a Dirac $\delta$-function as a solution.
We now do this for \erf{Langevin}, as an
aid the reader.

If $X$ is known at time $t$, then at that time
$P(x)=\delta(x-X)$. This then evolves
according to
\bqa
d\delta(x-X)&=&\delta(x-X-dX)-\delta(x-X)\\
&=&\delta[x-X-a(X)dt-\rt{D}dW(t)-edN(t)] \nl{-}
\delta(x-X).\label{delta1}
\eqa
We can treat the diffusion and jump noise in a unified fashion
by constructing a Taylor series of \erf{delta1} to
all orders of the first term on the right-hand side (RHS):
\bqa
d\delta(x-X)&=&\left(\exp\left\{\frac{\partial}{\partial
x}\left[-a(X)dt-\rt{D}dW(t) \right.\right.\right. \nn \\
&& \hspace{-2em}\left.\left.\phantom{\frac{1}{1}}\left.\phantom{\rt{1}}
 -edN(t)\right]\right\}  -1\right)\delta(x-X) .
\eqa
We now expand the exponential to all orders and use the \ito rules
$dN^{2}=dN$, $dW^{2}=dt$, with all other products being zero.
 This gives
\bqa
d\delta(x-X)
&=&\left\{-\frac{\partial}{\partial
x}\left[a(X)dt+\rt{D}dW(t)\right] \right. \nn \\ && +  \left.
\phantom{\frac{1}{1}}\frac{D}{2}\frac{\partial^{2}}{\partial
x^{2}}dt\right\}  \delta(x-X)
\nl{+}dN(t)\left[\delta(x-X-e)-\delta(x-X)\right] \nn \\
\eqa
Using the delta function to change $a(X)$ to $a(x)$ and then
averaging over the random variable $X$ gives the \ito equation for
the probability $P(x)\equiv {\rm E}[\delta(x-X)]$,
\bqa
dP(x)&=&\left\{\frac{\partial}{\partial x}
\left[-a(x)dt-\rt{D}dW(t)\right]+
\frac{D}{2}\frac{\partial^{2}}{\partial
x^{2}}dt\right\}\nl{\times}P(x) +dN(t)\left[P(x-e)-P(x)\right].
\label{CKE}
\eqa
The first ($dt$) term is deterministic drift, the second ($dW$) is
stochastic drift, the third ($dt$) is
diffusion and the last ($dN$) is the jump term.  We refer to this
equation as a stochastic differential Chapman-Kolmogorov equation
(SDCKE).  If there were no jump
term, then we would have a stochastic Fokker-Planck equation.

If averages
over $dW(t)$ and $dN(t)$ are taken then we have a standard
(deterministic)
differential Chapman-Kolmogorov equation
\bqa
\frac{\partial}{\partial t}P(x)&=&\left\{-\frac{\partial}{\partial x}
a(x)+\frac{D}{2}\frac{\partial^{2}}{\partial x^{2}}\right\}P(x) \nl{+}
g(x-e)P(x-e) - g(x)P(x),
\label{StandardCKE}
\eqa
where $g(x)$ enters from the definition
E$\sq{dN(t)f(X)}={\rm E}[g(X)f(X)]$.

\subsection{Kushner-Stratonovich Equation}

From the point of view of measurement, it useful to consider the
state of a classical system to be defined by the probability
distribution on {\em state space}.  This is the space of all relevant
physical quantities, called {\em state variables}.
In this paper the classical system will be a detector and the
associated electrical circuit.  Only one variable is necessary to
describe it so we can think of the distribution $P(x)$
that was introduced in the previous section as representing the state.

As the state of the system summarizes the observer's knowledge
it is logical that the state will change on the
basis of measurement upon the system.  From the {\em a priori} state 
$P(x)$
of the system and the result $r$ of measurement, the {\em a posteriori}
state $P(x|r)$ can be found using Bayesian inference
\cite{BayesBook} 
\beq
P(x|r)=\frac{P(r|x)P(x)}{P(r)}.
\label{BayesFirstRef}
\eeq
Here $P(x|r)$ reads as `the probability of $x$ given $r$'.  The
division by $P(r)$ is for normalization purposes, where
\beq
P(r)=\int dx\, P(r|x)P(x).
\eeq
Here $P(r|x)$ represents the probability of obtaining result $r$,
given that the state is $x$.  The expression in \erf{BayesFirstRef}
will be repeatedly used in this paper to obtain the
Kushner-Stratonovich equations (KSEs) for the classical system. This
is simply the continuous-in-time limit of Bayes theorem.
Rather than provide the general form of the KSE here, the reader will
be guided through the derivations of the KSEs (as they arise) from
fundamental considerations.

The unconditioned system state which we would use if the measurement had
been performed and ignored is given by
\beq
P'(x)=\int dr\, P(x|r)P(r),
\label{UnClass}
\eeq
where we have assumed that the results take on continuous values.
The prime is included to remind the reader that this is an {\em a
posteriori} state.  In the above we have assumed that there is no
 measurement back-action on the system  -- that is, the measurement  only
affects the observer's knowledge and not the dynamics of the system
itself -- so \erf{UnClass} reduces to $P(x)$ as one would
expect.  Of course, it is possible
to consider classical
measurements for which there is a measurement back-action.
This is also the case for all quantum measurements
 because of the \hei uncertainty principle.

\subsection{Stochastic Master Equation}

As noted in Sec.~\ref{Deriv}, the SME is a quantum version of the KSE,
including the evolution unconnected with the measurement as well. The
SME can be derived within the standard formalism of generalized
quantum measurement  (for simple introductions, see for example
Refs.~\cite{Wis95c} and \cite{WisDio01}). This formalism is simply the
quantum analogue of Bayes' theorem, generalized for
measurement back-action, as shown recently in Ref.~\cite{Har01}. In
this section we will not review the derivation of the SME from the
standard formalism, but will rather just quote the final results.

\subsubsection{Jumpy Trajectories}
\label{jumpTraj}
In many examples of continuous measurement on the output system, the
result can only take on two possible
values.  This is the case in ideal direct photodetection in which the
electromagnetic field is monitored with an ideal detection device that
responds to the presence of photons. In any infinitesimal time
interval a photon either is or is not detected. Such measurements give
rise to jumps in the conditional system state because the rare event
of a photon detection conveys to
the observer a finite amount of information about quantum
system. For example, if the fluorescent system were a TLA then a photon
detection
implies that the TLA now resides in the ground state. A null
result still causes a  non-unitary change to the
system state, but it is of infinitesimal magnitude. That is not to say that
it is unimportant, however, as over a finite length of time between
jumps, it will also cause a finite change in the system state.

Let the fluorescent system have lowering operator $c$, and assume
that the fluorescence can be formed into a beam of light. Also let an
optical local oscillator (that is, an effectively classical light field
from a laser) be added to the fluorescence before it is detected.
This can be done using a highly transmitting beam-splitter.  Let the
local oscillator (LO) amplitude added to the fluorescent beam be $\mu$. 
If we measure time in units
such that the photon flux from the system is $\an{c\dg c}$, then the
unobserved system obeys the (deterministic) master equation
\beq
\dot{\rho}(t)=-i[H,\rho(t)]+{\cal D}[c]\rho(t) \equiv {\cal L}\rho,
\label{MEFirst}
\eeq
where the superoperator ${\cal D}$ is defined, for arbitrary $B$, as
\bqa
{\cal D}[B]\rho&=&{\cal J}[B]\rho-{\cal A}[B]\rho\\
&\equiv&B\rho B^{\dag}-\half\left(B^{\dag}B\rho+\rho B^{\dag}B\right),
\eqa
where the two terms in the last line define ${\cal J}$ and ${\cal A}$
respectively. This master equation is of the Lindblad form
\cite{Lin76}, where we have assumed for simplicity
that the evolution apart from the
radiation damping term (${\cal D}[c]\rho$) can be described by a
Hamiltonian $H$.

To give an explicit description of the conditional system state evolution
under ideal photon detection, it is useful
to define the point process, $dN(t)$, that is equal to $1$
if there is a detection in the time interval $[t,t+dt)$ and  $0$
otherwise. Allowing for a detection efficiency $\eta$,
the statistics of $dN(t)$ are defined by
\bqa
dN(t)&=&dN(t)^{2},\label{dN2}\\
{\rm E}[dN(t)]&=&\eta{\rm Tr}[(c^{\dag}+\mu^{*})(c+\mu)\rho_{N}]dt.
\label{EdNeta}
\eqa
The subscript $N$ on the system state is to indicate that it is
conditioned on the detection events.
The evolution of the system state in terms of $dN(t)$ is given by
\cite{Wis94}
\bqa
d\rho_{N}&=&-\,dt{\cal
H}[iH+\half \eta c^{\dag}c+\eta\mu^{*}c+\half\eta|\mu|^{2}]\rho_{ N}
\nl{+}dN(t){\cal G}[\rt{\eta}(c+\mu)]\rho_{ N}
\nl{+}dt(1-\eta){\cal D}[c]\rho_{ N}.
\label{rhoNeta}
\eqa
This equation is a stochastic master equation (SME).
Note that we have dropped the time argument of the conditioned system
state, but maintained that of the increment, $dN(t)$.
This is done for convenience and, also, to emphasize the stochastic
nature of the detection events.  Time arguments are only included in
this paper when they will aid the reader's understanding.
The non-linear superoperators, ${\cal
G}$ and ${\cal H}$ are defined by
\bqa
{\cal G}[A]B &=&\frac{{\cal J}[A]B }{{\rm Tr}\left[{\cal
J}[A]B\right]}-B. \\
{\cal H}[A]B &\equiv& AB+BA\dg - {\rm Tr}[AB+BA\dg]B\label{calH}.
\eqa
It is possible to return to the ME (\erf{MEFirst}) by replacing $dN(t)$ in
\erf{rhoNeta} by its expectation value (\erf{EdNeta}).

\subsubsection{Diffusive Trajectories}
\label{DiffTraj}

In the limit of $\mu\raro\infty$ the rate of quantum jumps goes to
$\infty$ but the effect on the system of each one goes to zero.  This
leads to quantum trajectories of a diffusive nature, where the system
evolves continuously but non-differentiably in time.  In this limit
the photocounts give way to a photocurrent, $J$. Including a detection
efficiency $\eta$, and for a particular normalization, it
is given by
\beq
Jdt=\eta\an{ce^{-i\Phi}+c^{\dag}e^{i\Phi}}dt+\sqrt{\eta}dW(t),
\label{JGRVeta}
\eeq
where $dW(t)=\xi(t)dt$ represents Gaussian white noise \cite{Gar85} having the
properties
given in \erfs{xiMean}{xiSquare}. Here $\Phi = \arg(\mu)$, and
putting $\Phi=0$ in \erf{JGRVeta} corresponds to
measurement of the $x$ quadrature $c+c^{\dag}$, while $\Phi=\pi/2$
corresponds to measurement of the $y$ quadrature $-i(c-c^{\dag})$.
The system state evolution conditioned on this photocurrent is
\cite{Car93b,WisMil93a}
\beq
d\rho_{J}=dt{\cal
L}\rho_{J}+\left\{J(t)dt - {\rm E}[J(t)]dt\right\}{\cal H}[e^{-i\Phi}c]\rho_{J}
,
\label{rhoJeta}
\eeq
where again the subscript on  $\rho_{J}$ indicates that it is
conditioned on the recording of the current $J$.

\section{Realistic Photon Counting}
\label{PCounter}
An avalanche photodiode (APD) operating in Geiger mode produces a
macroscopic current pulse in response
to an incident photon.  This allows the observer to detect the
presence of photons and is the reason the device is also known as
a photon counter.
The size of the current pulse is large compared to
the sources of noise within the detection circuitry, which allows the
threshold current value for a detection to be set well above this noise.

\begin{figure}
\includegraphics[width=\col]{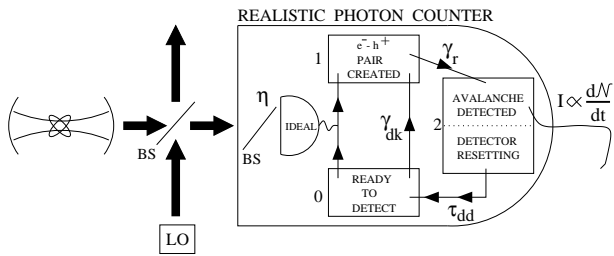}
\vspace{0.2cm}
\caption{Schematic diagram of the model used in this paper for a
realistic photon counter monitoring
a TLA placed in an optical cavity, the output of which
is combined with that of a weak local oscillator (LO) at a low
reflectivity beam
splitter (BS) before direct detection.
This will be the system on which investigation will focus in future numerical
work. In this
diagram we have used a BS of efficiency $\eta$ followed by an ideal
photon counter to represent the quantum efficiency of the realistic
detector.  The three states of the detector $0,1,2$ are explained in
the text.  Single arrow heads are used for Poisson processes.  The
photocurrent produced by the detector $I$ will consist of spikes at
the time of avalanches ($d{\cal N}=1$).}
\protect\label{PDDiag}
\end{figure}

\begin{figure}
\includegraphics[width=\col]{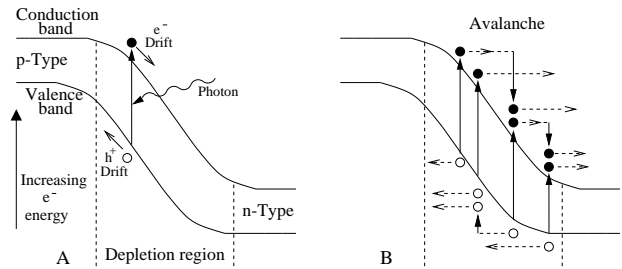}
\vspace{0.2cm}
\caption{Diagrams (A) and (B) show electron energy band diagrams for
a p-n junction operated under a beyond breakdown reverse bias. In
diagram (A) an incident photon causes the creation of an electron
(filled circles) - hole (circles) pair
that begins to separate due to the applied electric field.  The APD
would be in state $1$.
In (B)
the electron and hole collide with other charges leading to an
avalanche.  Whether the avalanche has reached threshold
or not determines if diagram (B) corresponds to the detector
being in state $2$ or $1$, respectively.}
\protect\label{PDStates}
\end{figure}

The APD essentially consists
of a p-n junction
 operated under a reverse bias greater than the breakdown voltage
\cite{OpFibV1,OpFibV3,OpFibTech}. Under these
conditions the diode  can be described  by just three classical
states \cite{OpFibV3} (see Fig.~\ref{PDDiag} and  Fig.~\ref{PDStates}).
This will
enable a mathematical model of the APD to be simply constructed.


The first state (0) is a stable low-current state in which
there are no charge carriers in the depletion region of the
junction. The transition from $0$ to the ``unstable'' intermediate state
($1$) takes place when
  an electron--hole ($e^{-}$--$h^{+}$)
pair is created in the depletion region
by an incident photon (with quantum efficiency $\eta$) or by
thermally initiated
`dark counts' occurring at a rate $\gamma_{{\rm dk}}$.
 Further
impact ionization, under the influence of the electric
field, leads to an avalanche forming.  At this stage (before the
avalanche has been detected) the APD is still
in state $1$. The avalanche continues to build until the current
reaches some threshold value and a detection is registered, thus changing
the state of the APD to $2$.  Due to the stochastic manner in
which the avalanche spreads, the dwell time in state $1$
is not deterministic. We model the $1$ to $2$ transition as a
Poissonian process with rate $\gamma_{\rm r}$. We call the mean dwell
time, $\gamma_{\rm r}^{-1}$, the `response time'.
The avalanche, once detected, is arrested by the application of a
negative-going voltage pulse that temporarily brings the bias voltage below the
breakdown value \cite{OpFibTech}.  This results in a fixed
 `dead time', $\tau_{{\rm dd}}$, during which the APD cannot
detect photons, after which it is restored to state 0.

\subsection{Physical Explanation of Parameters}

Now that our basic model for the APD has been introduced, a further
physical explanation is given for the
quantum efficiency, response time, the dead time and dark counts.

The quantum efficiency, $\eta$, is largely determined by two
considerations.  These are the fraction of incident photons that are
absorbed in the active region of the junction and the fraction of
$e^{-}$--$h^{+}$ pairs that lead to an avalanche \cite{OpFibV3}.  The first is
determined by the absorption co-efficient of the material at the
wavelength of interest and, also, the thickness of the depletion
region.  If the absorption co-efficient is too large then the photons
will be absorbed near the surface, where there exists only a weak
biasing electric field.  Too small a co-efficient allows the
light to pass straight through.  The second consideration is a strong
function of the overvoltage (beyond breakdown), with higher fields ensuring
that
the creation of a charged pair leads to an avalanche.  Unfortunately,
this also leads to approximately exponentially increased dark counts
\cite{etaRecord}.

The response rate, $\gamma_{{\rm r}}$, is also a strong function of
the overvoltage due to its effect on the rate at which charge
carriers move through the junction.
  Other important considerations are the dimensions of the
junction, which alter the time taken for the avalanche to spread
throughout.  Two important mechanisms that facilitate the spreading
are the direct electrical diffusion of carriers into adjacent regions
of the junction
and the emission of photons from hot carriers that are absorbed in a
different region \cite{Lacita}.

The dead time, $\tau_{{\rm dd}}$, is generally quite long as all
charge carriers must be swept out of the active region before the
overvoltage can be re-applied without another avalanche resulting.
The application of a negative-going voltage pulse (active quenching)
is more effective than letting a resistance in series with the
junction reduce the reverse bias below breakdown as the current
increases (passive quenching).  The active quenching of the avalanche
can be initiated as soon as soon as the current reaches a threshold
value.  Minimizing the circuit noise allows the threshold to be
lowered thus reducing the response time and,
also, the duration of the voltage pulse.

The dark count rate, $\gamma_{{\rm dk}}$, can be reduced to
insignificance with the use of a high quality Silicon APD
\cite{OpFibV3}.  However,
such devices are not appropriate for detection at all wavelengths.  In
other materials the existence of free-carrier trapping centers and
thermally generated carriers creates a considerable problem.  During
an avalanche, carriers can become trapped in a region of the junction,
only to be released after the above breakdown
voltage has been restored (causing an afterpulse).
This leads to a dark count which is signal-dependent as the
number of trapping centers generally increases with the frequency of
avalanches.
Cooling will reduce the thermal dark count.
These complexities are not introduced into the APD model in this paper.

\subsection{Realistic Quantum Trajectories}
\label{QTrajPhotonCounter}
Our aim is to derive the
quantum trajectories for the quantum system (the source of the light
entering the APD) {\em conditioned on the
observation of an avalanche}.  In order that the derived equations
be applicable to different measurement schemes, we consider the
output field of the system of interest to be combined at a low
reflectivity beam
splitter with that of a
weak local oscillator (LO), having transmitted amplitude $\mu$.  If
the LO is not weak, and contributes a large photon flux, then a
photon counter is not the appropriate type of detector.  On the other
hand, $\mu$ can be set to zero to obtain direct detection without
problems.

In Fig.~\ref{Method} the general method of deriving realistic quantum
trajectories was given.  The point that some of the steps in the
method are interchangeable was made in the caption to that figure.
For the APD, we will in fact perform
the conditioning upon measurement after the
unobserved processes are averaged over.  This will keep the
equations that we manipulate during the derivation as simple as possible.

\subsubsection{Stochastic Master Equation}
As described in the introduction, the detection device can be
treated in a classical manner.  This implies that the presence of
an observer monitoring the detailed behaviour of the charge-carriers
in the p-n junction would not alter the detection results.  The
hypothetical observer could also, in principle, deduce the incidence
of photons that
were absorbed outside the active region of the junction.  In this way,
a list of the times at which photons arrived could be compiled.  Of
course, such sensitive measurements are currently out of reach due to
technological reasons, but the point is that it is consistent with
quantum measurement theory to assume a quantum jump unraveling of the
ME.  In fact, we even assume at this stage that the measurement can
have unit efficiency, as it is conceptually more satisfying to
explicitly include inefficiency in our
description of the APD. Thus the fictitious conditional dynamics of
the system are those of \erf{rhoNeta}, with $\eta=1$.

\subsubsection{Stochastic Differential Chapman-Kolmogorov Equation}
We can now formulate equations of motion for the probability of each of the
three
classical states of the APD being occupied.  Because the state space
is discrete, averaging over the stochastic processes would necessarily
yield a classical master equation \cite{Gar85} and we could refer
to the equations we derive in this section as a stochastic classical master
equation. However, to solidify 
for the reader the processes involved in forming a realistic quantum 
trajectory (described in Sec.~\ref{Deriv}), we refer to the stochastic 
probability equations as SDCKEs. The
stochastic equation would apply if all transitions were monitored,
and that when an
electron-hole pair is created a distinction can be drawn between
those created via photon absorption and those via thermal fluctuations (dark
counts).  Because of the classical nature of the APD dynamics, this
is logically possible. Eventually we will average over
processes that a realistic observer cannot monitor.

There are a number of processes that contribute to the evolution of
the occupation probabilities.  Here, we will treat dark counts, avalanches,
APD resetting and incident photons from the monitored quantum system.
Since all transitions are being monitored by a fictitious observer
and we assume, for
simplicity, that the initial state of the APD is known the evolution
of the probabilities is described by jump terms only.  If we use
$dN_{{\rm dk}}(t)=0,1$ to define the occurrence of
transitions due to dark counts, $dN'(t)$
to indicate transitions due to photons from the quantum system of
interest and $d{\cal N}'(t)$ for avalanches then
\bqa
dP_{0}&=&-dN_{{\rm dk}}(t)-dN'(t)+d{\cal N}'(t-\tau_{{\rm
dd}})\label{Prob0}\\
dP_{1}&=&dN_{{\rm dk}}(t)+dN'(t)-d{\cal N}'(t)\label{Prob1}\\
dP_{2}&=&d{\cal N}'(t)-d{\cal N}'(t-\tau_{{\rm dd}}).\label{Prob2}
\eqa
From these SDCKEs it can be seen that photons and dark counts,
which lead to the creation of an
electron-hole pair, take the detector from state $0$ to 
state $1$.  Avalanches cause the transition $1\raro 2$ and the
resetting of the detector, with the use of the delayed time
$t-\tau_{{\rm dd}}$, returns it to the ready state $0$.

The use of the prime on $dN'(t)$ indicates that this point process is
associated with, but not the same as, $dN(t)$ which appears in 
\erf{rhoNeta}.  In fact
$dN'(t)$ is equal to $dN(t)$, except for having some of the ones removed.
This leads to
\beq
dN(t)dN'(t)=dN'(t).
\label{lack}
\eeq
Their expectation values are related by
\beq
{\rm E}[dN'(t)]=\eta P_{0}{\rm E}[dN(t)].
\label{comp}
\eeq
This reflects the two ways in which it is possible to have a ``missed 
detection'' [$dN(t)=1,\;dN'(t)=0$], namely,  
inefficiency or the APD's not being in the ready state.

The use of the prime on $d{\cal N}'(t)$ is to distinguish an
avalanche from an observed avalanche, denoted by $d{\cal N}(t)$.  We
note that such a distinction could be necessary if the efficiency of
avalanche detection was less than unity.  Although we do not include
an avalanche detection inefficiency here we feel that the prime may aid the
reader in understanding the need for the conditioning of the equations
upon measurement of the presence of avalanches, which will occur
later.
The	statistics of $d{\cal N}'(t)$ are defined by its expectation value
\beq
{\rm E}[d{\cal N}'(t)]=\gamma_{{\rm r}}P_{1}dt.
\eeq
This comes from our assumption that the avalanche's reaching its 
threshold value is a Poisson process. 
Finally, the expectation value of $dN_{{\rm dk}}(t)$ is
\beq
{\rm E}[dN_{{\rm dk}}(t)]=\gamma_{{\rm dk}}P_{0}dt,
\label{dkexp}
\eeq
reflecting another Poisson assumption, here for the dark counts.

\subsubsection{Joint Stochastic Equations}

The evolution of the quantum system is linked to that of the detector
through $dN'(t)$.  To obtain realistic quantum trajectories for the
supersystem, which includes the quantum system and the classical
detector states, we form the quantity
\beq
\rho_{ i}=P_{ i}\rho_{N},
\label{form}
\eeq
where $\rho_{ i}$ is the unnormalized state of the system
given that the detector is in state $i$.
The normalized system state is
\beq
\rho_{}=\rho_{0}+\rho_{1}+\rho_{2}.
\label{TrOverAPD}
\eeq
It is only possible to
assume that the quantum and detector states factorize if there are no
correlations between them.  This will be the case until we average 
over unobserved processes.  Thus the evolution of $\rho_{
i}$ can be found from
\beq
\rho_{ i}+d\rho_{ i}=(P_{ i}+dP_{i})(\rho_{N}
+d\rho_{N}),
\label{rhoi}
\eeq
with the use of the increments in \erf{rhoNeta} and
\erfs{Prob0}{Prob2}.  Conditioning upon the measurement of $d{\cal
N}(t)$ will be done as a later step.
Substitution into \erf{rhoi} with the use of \erf{lack} gives
\begin{widetext}
\bqa
\rho_{0}(t+dt)&=&\rho_{0}+P_{0}\left[dN(t){\cal G}[c+\mu]-dt{\cal
H}[iH+\half c^{\dag}c+\mu^{*}c+\half|\mu|^{2}]\right]\rho_{N}\nonumber\\
&&-\left[dN_{{\rm dk}}(t)+dN'(t)-d{\cal N}'(t-\tau_{{\rm
dd}})+dN'(t){\cal G}[c+\mu]\right]\rho_{N}\label{one}\\
\rho_{1}(t+dt)&=&\rho_{1}+P_{1}\left[dN(t){\cal G}[c+\mu]-dt{\cal
H}[iH+\half c^{\dag}c+\mu^{*}c+\half|\mu|^{2}]\right]\rho_{N}\nonumber\\
&&+\left[dN_{{\rm dk}}(t)+dN'(t)-
d{\cal N}'(t)+dN'(t){\cal G}[c+\mu]\right]\rho_{N}\\
\rho_{2}(t+dt)&=&\rho_{2}+P_{2}\left[dN(t){\cal G}[c+\mu]-dt{\cal
H}[iH+\half c^{\dag}c+\mu^{*}c+\half|\mu|^{2}]\right]\rho_{N}\nonumber\\
&&+\left[d{\cal N}'(t)-d{\cal N}'\left(t-\tau_{{\rm
dd}}\right)\right]\rho_{N},\label{three}
\eqa
\end{widetext}
remembering that all time arguments not stated are $t$.
The linking of the quantum system and the APD is manifest in the
$dN'(t){\cal G}[c+\mu]\rho_{N}$ terms.  If we were now to average
over the emissions from the quantum system a correlation would still
develop between the quantum state and the APD.

\subsubsection{Average Over Unobserved Processes}

In order to obtain trajectories that are based on information that is
realistically available in the laboratory, we must now average over
unobserved processes.  These are the photon emissions and dark
counts.  Strictly speaking, we should also
average over $d{\cal N}'(t)$ since $d{\cal N}(t)$ was stated as
specifying observed avalanches. However, we have not yet performed the
conditioning due to the detection of the avalanches so these
transitions will be left for the moment.  Doing the averages before
the measurement does not change the final result, but does make the
equations more simple.

The relevant averages are \erf{EdNeta}, \erf{comp} and \erf{dkexp}.
  Substitution of these into
\erfs{one}{three} gives
\bqa
\rho_{0}(t+dt)&=&\rho_{0}+dt\left({\cal L}-\gamma_{{\rm dk}}-\eta{\cal
J}[c+\mu]\right)\rho_{0}\nl{+}d{\cal N}'(t-\tau_{{\rm
dd}})\rho\label{0Ave} \\
\rho_{1}(t+dt)&=&\rho_{1}+dt{\cal L}\rho_{1}+
dt\left(\gamma_{{\rm dk}}+\eta{\cal J}[c+\mu]\right)\rho_{0}
\nl{-}d{\cal N}'(t)\rho\\
\rho_{2}(t+dt)&=&\rho_{2}+dt{\cal L}\rho_{2}
+d{\cal N}'(t)\rho - d{\cal N}'\left(t-\tau_{{\rm dd}}\right)\rho.
\nn \\
\label{2Ave}
\eqa
Note that $\rho$ is given by \erf{TrOverAPD} and ${\cal L}$ is 
specified in \erf{MEFirst}.

\subsubsection{Kushner-Stratonovich Equation}
The probabilities assigned to the occupation of the three APD states
by the realistic observer will undergo jumps in the event of an
avalanche.  If there is no avalanche then this also provides
information to the observer about the probabilities.  Let us first
consider the effect of an avalanche.  The new states will be given by
\beq
\rho_{i|d{\cal N}(t)=1}=\frac{\rho_{i}}{{\rm Tr}[\rho_{i}]}P_{i|d{\cal
N}(t)=1}\;,
\eeq
where $P_{i|d{\cal N}(t)=1}$ is the probability of the APD being in
state $i$ given that an avalanche was observed.  These probabilities can
be simply calculated using Bayes conditional probability theorem
\cite{BayesBook}
\beq
P_{i|d{\cal N}(t)=1}=\frac{P_{d{\cal N}(t)=1|i}P_{i}}{P_{d{\cal N}(t)=1}}.
\label{bayesdk}
\eeq
Here, $P_{d{\cal N}(t)=1|i}$ is the probability of observing an
avalanche given that the APD is in state $i$, $P_{i}$ is the prior
occupation of state $i$ and $P_{d{\cal N}(t)=1}$ is the probability of
observing an avalanche.  The relevant probabilities are
\bqa
P_{d{\cal N}(t)=1|0}=P_{d{\cal N}(t)=1|2}&=&0\\
P_{d{\cal N}(t)=1|1}&=&\gamr dt\\
P_{d{\cal N}(t)=1}&=&\gamma_{{\rm r}}P_{1}dt.
\eqa
This leads to the following three results for the states
\bqa
\rho_{0|d{\cal N}(t)=1}&=&0\\
\rho_{1|d{\cal N}(t)=1}&=&\frac{\rho_{1}}{{\rm Tr}[\rho_{1}]}\\
\rho_{2|d{\cal N}(t)=1}&=&0.
\eqa
If there is no observed avalanche then we need to consider
probabilities of the form $P_{i|d{\cal N}(t)=0}$. This can be
calculated from
\beq
P_{i|d{\cal N}(t)=0}=\frac{P_{d{\cal N}(t)=0|i}P_{i}}{P_{d{\cal N}(t)=0}},
\label{bayesdk2}
\eeq
with the probabilities interpreted in the usual fashion.
The relevant expressions are
\bqa
P_{d{\cal N}(t)=0|0}=P_{d{\cal N}(t)=0|2}&=&1\\
P_{d{\cal N}(t)=0|1}&=&1-\gamma_{{\rm r}}dt\\
P_{d{\cal N}(t)=0}&=&1-\gamma_{{\rm r}}P_{1}dt.
\eqa
The new states based on these probabilities are, to first order in
$dt$,
\bqa
\rho_{0|d{\cal N}(t)=0}&=&(1+\gamma_{{\rm r}}P_{1}dt)\rho_{0}\\
\rho_{1|d{\cal N}(t)=0}&=&(1-\gamma_{{\rm r}}dt+
\gamma_{{\rm r}}P_{1}dt)\rho_{1}\\
\rho_{2|d{\cal N}(t)=0}&=&(1+\gamma_{{\rm r}}P_{1}dt)\rho_{2}.
\eqa
We can summarize the avalanche and no avalanche conditioning with
\bqa
\rho_{0|d{\cal N}(t)}&=&[1+\gamma_{{\rm r}}P_{1}dt-d{\cal
N}(t)]\rho_{0}\label{cond0}\\
\rho_{1|d{\cal N}(t)}&=&\left(1-\gamma_{{\rm r}}dt+
\gamma_{{\rm r}}P_{1}dt\right)\rho_{1}\nl{+}d{\cal N}(t)\left(\frac{1}{{\rm
Tr}[\rho_{1}]}-1\right)\rho_{1}\label{cond1}\\
\rho_{2|d{\cal N}(t)}&=&[1+\gamma_{{\rm r}}P_{1}dt-d{\cal
N}(t)]\rho_{2}.
\label{cond2}
\eqa
Although we have already formed joint stochastic equations and 
averaged over unobserved processes, the above equations represent the step 
in Fig.~\ref{Method} where we obtain the KSEs for classical probabilities of 
the APD states.

\subsubsection{Superoperator Kushner-Stratonovich Equation}

The expressions in \erfs{cond0}{cond2} can now be substituted into the RHS of
\erfs{0Ave}{2Ave}, using \erf{TrOverAPD} for $\rho$.   However, a
simplifying deduction is that when the APD resets, the state of the
quantum system must be equal to $\rho_{2}$, which will already be
normalized.  We can therefore replace $\rho$ by $\rho_{2}$ as the
multiplier of the $d{\cal N}'(t-\tau_{{\rm dd}})$ term.
Similar reasoning allows us to ignore terms such as
 $-d{\cal N}(t)\rho_{2}$ and $\gamma_{r}{\rm
Tr}[\rho_{1}]\rho_{2}dt$ that will appear in the equation for $\rho_{2}$.
  The assumption of a known initial APD state allows us to make the
following statement:  {\em at all times either
$P_{0}+P_{1}=1$ or $P_{2}=1$}.  After an avalanche, the detector is
known to be in state $2$ for a time $\tau_{{\rm dd}}$.  Once the APD
has been reset a realistic observer will not know if the APD is in
state $0$ or $1$ (but will know that it is not in state $2$)
until the time of the next avalanche.

Also
note that $d{\cal N}(t)d{\cal N}'(t)=d{\cal N}(t)=d{\cal N}'(t)$ since here
we are
assuming that all avalanches are detected by the observer.  This
allows us to drop the prime on the non-Markovian resetting terms
$d{\cal N}'(t-\tau_{{\rm dd}})$, rather than perform a repetitive
conditioning on the observation of the resetting process, which is
actually just inferred from the observation of an avalanche.

Making these simplifications after substitution into \erfs{0Ave}{2Ave} gives
\bqa
d\rho_{0}&=&dt\left({\cal L}-\gamma_{{\rm dk}}-\eta{\cal
J}[c+\mu]+\gamma_{{\rm r}}{\rm Tr}[\rho_{1}]\right)\rho_{0}
\nl{-}d{\cal N}(t)\rho_{0}+d{\cal N}(t-\tau_{{\rm
dd}})\rho_{2}\label{dp0}\\
d\rho_{1}&=&dt\left[\left({\cal L}-\gamma_{{\rm
r}}+\gamma_{{\rm r}}{\rm Tr}[\rho_{1}]\right)\rho_{1}+
\left(\eta{\cal J}[c+\mu]+\gamma_{{\rm dk}}\right)\rho_{0}\right]
\nl{-}d{\cal N}(t)\rho_{1}\label{dp1}\\
d\rho_{2}&=&dt{\cal L}\rho_{2}+d{\cal
N}\left(t\right)\frac{\rho_{1}}{{\rm Tr}[\rho_{1}]}
-d{\cal N}\left(t-\tau_{{\rm dd}}\right)\rho_{2}. \label{dp2}
\eqa
 The above equations are the realistic
quantum trajectories based on the observation of avalanches in the APD. We
term the equations SKSEs as
we have obtained a quantum analog of the KSE in that from measurement 
we are conditioning the state of a
supersystem that contains a quantum system, as explained in Sec.~\ref{Deriv}.
The conditioned system state is
\beq
\rho_{\cal N} = \rho_{0}+\rho_{1}+\rho_{2}.
\eeq
Note that \erfs{dp0}{dp2} are nonlinear, so that the state remains
normalized. This is in contrast to the corresponding
equations in \cite{WarWisMab01}, which were linear and generated
unnormalized states. For simulation purposes, it is more convenient to
use the unnormalized forms.

\subsection{Ignoring Imperfections}

Depending on the experimental situation, one or more of the
imperfections in the APD may be negligible. Easy limits to take are those
of $\eta =1$ and $\gamma_{{\rm dk}}=0$, as these values can be
substituted into the already derived equations.
More involved limits are those where $\gamma_{{\rm r}}^{-1}$  and
$\tau_{{\rm dd}}$ become small.  These imply that one or two of the detector
states are superfluous. The first  limit can be treated by a process of
adiabatic elimination.

When $\gamma_{{\rm r}}$ is much larger than
the system rates (given by the eigenvalues of ${\cal L}$) we can
adiabatically eliminate state $1$ of the detector.  When
$\rho_{1}$ responds quickly to changes,
$\dot{\rho}_{1}$ in \erf{dp1} can be set to zero to give the
slaved value
\beq
\rho_{1}=\frac{1}{\gamma_{{\rm r}}}(\eta{\cal
J}[c+\mu]+\gamma_{{\rm dk}})\rho_{0}.
\label{rho1slave}
\eeq
Note that ${\cal L}\rho_{1}$ and $\gamma_{{\rm r}}{\rm
Tr}[\rho_{1}]\rho_{1}$ have been ignored as compared to
$\gamma_{{\rm r}}\rho_{1}$ and that \erf{rho1slave} does not
require the inclusion of a jump term due to avalanches (as these are
inherited from the dependence upon $\rho_{0}$).
As $\rho_{1}$ can now be determined
from \erf{rho1slave}, the number of differential equations for real
variables that need to be evolved is greatly reduced

If $\gamma_{{\rm r}}\rightarrow\infty$ then state $1$
can be ignored completely, with \erfs{dp0}{dp2} reducing to
\bqa
d\rho_{0}&=&dt\left({\cal
L}-\gamma_{{\rm dk}}-\eta{\cal J}[c+\mu]+
{\rm E}[d{\cal N}(t)]\right)\rho_{0}
\nl{-}d{\cal N}(t)\rho_{0}+d{\cal N}(t-\dtime)\rho_{2},
\label{dp0No1}\\
d\rho_{2}&=&dt{\cal
L}\rho_{2}-d{\cal N}(t-\dtime)\rho_{2}\nl{+}d{\cal N}(t)\frac{(\eta{\cal
J}[c+\mu]+\gamma_{{\rm dk}})\rho_{0}}{{\rm Tr}[\left(\eta{\cal
J}[c+\mu]+\gamma_{{\rm dk}}\right)\rho_{0}]}.
\label{dp2No1}
\eqa
These equations describe the detector jumping straight from the ready
state to the dead state.  Thus, the exact time at which a charged
$e^{-}$--$h^{+}$ pair is created is known, but creation by
a dark count or a
photon is still indistinguishable.  The statistics of $d{\cal N}(t)$
 are now defined by $
{\rm E}[d{\cal N}(t)]=dt{\rm Tr}[(\eta{\cal J}[c+\mu]+\gamma_{{\rm
dk}})\rho_{0}]$ and we have $\rho_{\cal N}=\rho_{0}+\rho_{2}$.  Note
that only one of $\rho_{0}$ and $\rho_{2}$ will contribute to
$\rho_{\cal N}$ at any particular time.  If the expression for
${\rm E}[d{\cal N}(t)]$ were included in \erf{dp0No1} then the
$-dt\gamma_{{\rm dk}}\rho_{0}$ term would cancel due to
it only having an effect on the transition rate out of state $0$,
which is now effectively being monitored.

If, on the other hand, the dead time goes to zero, then state $2$ of the
detector can be
removed.  Thus, when an avalanche occurs, the detector jumps straight
to the ready state.  The relevant equations are then (with state $1$
not adiabatically eliminated)
\bqa
d\rho_{0}&=&dt\left({\cal
L}-\gamma_{{\rm dk}}-\eta{\cal J}[c+\mu]+\gamma_{{\rm r}}{\rm
Tr}[\rho_{1}]\right)\rho_{0}
\nl{-}d{\cal N}(t)\left(\rho_{0}-\frac{\rho_{1}}{{\rm Tr}[\rho_{1}]}\right)
\label{dp0No2},\\
d\rho_{1}&=&dt\left[\left({\cal
L}-\gamma_{{\rm
r}}+\gamma_{{\rm r}}{\rm Tr}[\rho_{1}]\right)\rho_{1}+
\left(\eta{\cal J}[c+\mu]
+\gamma_{{\rm dk}}\right)\rho_{0}\right]
\nl{-}d{\cal N}(t)\rho_{1}.
\label{dp1No2}
\eqa
Here, the statistics of $d{\cal N}(t)$ are defined as for \erfs{dp0}{dp2}
but we now have $\rho_{\cal N}=\rho_{0}+\rho_{1}$.

Finally, we consider $\gamma_{{\rm r}} \to \infty$ together with
$\tau_{{\rm dd}}=0$.  This leaves an equation for $\rho_{0}$
only.  It is,
\bqa
d\rho_{0}&=&dt\left({\cal
L}-\gamma_{{\rm dk}}-\eta{\cal J}[c+\mu]+
{\rm E}[d{\cal N}(t)]\right)\rho_{0} + d{\cal
N}(t)\nl{\times}\left(\frac{(\eta{\cal
J}[c+\mu]+\gamma_{{\rm dk}})}{{\rm Tr}[\left(\eta{\cal
J}[c+\mu]+\gamma_{{\rm dk}}\right)\rho_{0}]}-1\right)\rho_{0},
\label{dp0No12}
\eqa
with ${\rm E}[d{\cal N}(t)]$ defined as for \erfs{dp0No1}{dp2No1}.
If $\gamma_{{\rm dk}}=0$ then \erf{dp0No12} reduces to
\erf{rhoNeta}. Obviously, here we have $\rho_{\cal N}=\rho_{0}$.

\section{Photoreceiver}
\label{Photoreceiver}

\begin{figure}
\includegraphics[width=\col]{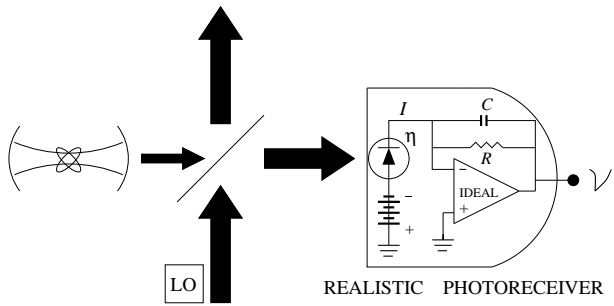}
\vspace{0.2cm}
\caption{Homodyne detection by a realistic photoreceiver
of a TLA placed in an optical cavity.  The photocurrent generated by a below
breakdown p-i-n junction, of efficiency $\eta$,
is input to a transimpedance amplifier.  The voltage measured in the
laboratory is ${\cal V}$.}
\protect\label{PRDiag}
\end{figure}

When the incident photon flux is high, as in
homodyne detection, an avalanche photodiode cannot be used because of
the long dead time. Instead, a p-i-n photodiode is more appropriate.
The inclusion of the
high resistivity intrinsic region, i, allows the widening of the
depletion region, which facilitates the absorption of light
\cite{OpFibV3}.  Because there is no avalanche, an external amplifier,
such as a transimpedance
amplifier (see Fig.~\ref{PRDiag}) must be used
\cite{OpFibTech}. We consider the whole system of photodiode plus
amplifier to constitute the photoreceiver.

When a photon strikes the depletion region of the p-i-n
junction, an electron--hole pair (that does not recombine)
 is produced with probability equal to the quantum
efficiency $\eta$. The charge carriers drift under the influence of the
below-breakdown reverse bias, and the resultant
 current $I$ is fed into an operational amplifier (op-amp) set up as a
 transimpedance amplifier. This has a low effective input impedance,
 so that the diode acts as a current source, and $I$ is converted into
  a voltage drop $V$ across the
feedback resistor, $R$.  The
capacitor $C$, in parallel with $R$, represents the total
capacitance from the output of the op-amp back to its input,
including capacitance added deliberately for the smoothing of noise
and oscillations.  If no electronic noise were present, the output
voltage of the photoreceiver would be a filtered version of the input
signal given, in the frequency domain, by
\beq
V(\omega )=\frac{-I(\omega)R}{1+i\omega RC},
\label{Vfreq1}
\eeq
where the negative is included because the output of the op-amp
will be such that the input is kept at virtual ground.

It should be noted that if this were the case (that is, if there were
no noise) then
the input $I$ could be perfectly reconstructed from the filtered
signal $V$, since \erf{Vfreq1} is equivalent to
\beq \label{IVse}
I(t) = -C\dot{V}(t)-\frac{V(t)}{R}.
\eeq
Thus the resultant quantum trajectories would be no different
from those of a photoreceiver with infinite bandwidth, since the input
current to the amplifier is determinable. Everything of
interest results therefore from the presence of excess noise.
We include only the Johnson noise $V_{{\rm J}}$ from the
feedback
resistor, which has a flat spectrum $S_{{\rm J}}=4k_{{\rm B}}TR$.  Here,
$k_{{\rm B}}$ is the Boltzmann constant and $T$ is the temperature.
This simplification (neglecting contributions from voltage noise of the
operational amplifier and dark counts from the photodiode)
 can be justified for practical receivers with $R\sim 10$k$\Omega$
 \cite{MabPriv}.

The output voltage ${\cal V}$ from the photoreceiver is given by sum
 of the filtered signal and the Johnson noise
\beq {\cal V}=V+V_{{\rm J}}.
\label{Vo}
\eeq
Our aim is to find the quantum trajectory for the system, conditioned
on continuously monitoring ${\cal V}$. Since the voltage $V$, which
describes the detector state,
is a continuous variable, in this case $\mathbb{S}= \mathbb{R}$, the real line,
and the supersystem can be described by an operator function
$\rho(V)$. Finding the stochastic equation of motion for $\rho(V)$ is quite
involved.

\subsection{Realistic Quantum Trajectories}
\label{RQTraj}
In the following derivation we follow the order schematically
illustrated in Fig.~\ref{Method} more closely than in the case of the
APD.  The averaging over the unobserved processes is performed as a
final step.

We begin by taking the output current $I$ of the photodiode to be that from a
perfect (apart from its efficiency $\eta$) unbalanced
homodyne detection
of the output field of the system.
For a LO tuned to the resonant frequency of the system
$\omega_{0}$,  of power ${\rm P}$, and
phase $\Phi$, the current is \cite{Car93b,WisMil93a}
 \beq
 I=e\sqrt{{\rm P}/\hbar\omega_{0}}\left[\eta\langle e^{-i\Phi}c
+e^{i\Phi}c^{\dag}\rangle +\rt{\eta}\xi(t)\right],
\label{I}
\eeq
where we have ignored the DC component due to the LO power and $e$ is
the electron charge. This current (in amps) is just the scaled current $J(t)$
introduced in \erf{JGRVeta}, multiplied by $e\sqrt{{\rm P}/\hbar\omega_{0}}$.

The Gaussian white noise \cite{Gar85} in \erf{I}
can be considered to arise
from two independent sources: the Poisson statistics of the
LO and the vacuum noise introduced by the
inefficiency of the photodiode. The distinction between
noise from
these two sources will be illustrative when numeric simulations 
of the trajectories are performed (in a later paper). 
For now we need only consider $\xi(t)$.

\subsubsection{Stochastic Master Equation}

The evolution of the quantum system conditioned on $I$ is given, in terms of
the noise $\xi(t)$, by \erf{rhoJeta} and is restated here for the
reader's convenience
\beq
d\rho_{I}=dt\left\{{\cal
L}+\rt{\eta}\xi(t){\cal H}[e^{-i\Phi}c]\right\}\rho.
 \label{rhoI}
 \eeq
The subscript $J$ in \erf{rhoJeta} has been changed to $I$.
The fact that $\rho$ on the RHS of \erf{rhoI} would usually be
conditioned on previous values of $I(t)$
is to be understood. A realistic observer does not have direct access
to $I(t)$, so later we will average over it.

\subsubsection{Stochastic Differential Chapman-Kolmogorov Equation}

Now, \erf{Vfreq1} is equivalent to the
stochastic equation (\ref{IVse}).
Since the voltage $V$ is not directly measured, we must consider a
distribution $P(V)$ for it. Assuming that $C>0$, and, for the moment,
 that $I$ is known, \erf{IVse} can be converted to
an \ito stochastic Fokker-Planck equation
for $P(V)$ conditioned on the photocurrent.  This is done using the
theory of Sect.~\ref{Trajrefer}, with the result being
\beq dP_{I}(V)=\left(\frac{\partial}{\partial
V}\frac{V+IR} {RC}+\frac{{\rm P}\eta
e^{2}}{2\hbar\omega_{0}C^{2}}\frac{\partial^{2}} {\partial
V^{2}}\right)P(V)dt.
\label{PI}
\eeq
A Fokker-Planck equation is one specific type of the broader class
of differential Chapman-Kolmogorov equations.
As in \erf{rhoI} we are using the convention that subscripts
indicate that the increment is conditioned
on that result.  For example, the probability distribution for $V$
given a current $I$ is $P_{I}(V)\equiv P(V|I)$.  It
should be noted that \erf{PI} is different from a standard
Fokker-Planck equation in that $I$ still contains a white noise term.
It is essential to retain this term as it is the same noise that
appears in \erf{I} - this will be used later in our derivation to link
the evolution of the quantum system with that of the circuit of the
photoreceiver.

\subsubsection{Kushner-Stratonovich Equation}

Next we need to determine the effect of the measurement of ${\cal V}$
on $P(V)$.  This can be
calculated by using Bayes' theorem
\beq
P_{{\cal V}}(V)=\frac{P_{V}({\cal V})P(V)}{P({\cal V})}.
\label{bayes}
\eeq
Remembering that the
Johnson noise is white,
 it follows from \erf{Vo} that
$P_{V}({\cal V})$ is a Gaussian with mean $V$ and variance $4k_{{\rm
B}}TR/dt$.  This enables us to calculate the properties of $P({\cal
V})$ from
\bqa
P({\cal V})&=&\int dVP_{V}({\cal V})P(V)
\label{1stRepeat}\\
&=&({2\pi\beta})^{-1/2}\int dV \exp(-({\cal
V}-V)^{2}/2\beta)P(V)\\
&=&\int dV\exp\left[(V-\langle V\rangle)(2{\cal V}-V-\langle V\rangle )
/2\beta\right] P(V) \nl{\times}
({2\pi\beta})^{-1/2}{\exp(-({\cal
V}-\an{V})^{2}/2\beta)}.
\label{Factor}
\eqa
For convenience, the variance $4k_{{\rm B}}TR/dt$ in the above
equations has been rendered by $\beta$. For an
expansion of the exponential inside the integral to be possible the
factorization of \erf{Factor} had to be made.  The variance in $V$
is much smaller than that of
${\cal V}$, which is of order $O(1/dt)$, due to the presence of
the white Johnson noise.

Expanding terms in the above expression to
leading order, we find eventually that
\beq
P({\cal V})=\frac{\exp(-({\cal
V}-\an{V})^{2}/2\beta)}{\sqrt{2\pi\beta}}\left[1+O\left(dt^{3/2}\right)\right].
\eeq
It follows that we can write, correct to $O(1)$,
\beq
{\cal V}=\langle V\rangle+\rt{4k_{{\rm B}}TR}\frac{\dwj}{dt},
\label{defdWJ}
\eeq
where $\dwj/dt$ is another Gaussian white noise source,
independent of $\xi(t)$.  Despite the fact that it arises from the
Johnson noise, it is not the actual Johnson noise.  It represents
noise that the observer can determine from the measurement result and
$\an{V}$.  Labeling the actual
Johnson noise by $\dwjp$ we have
\beq
{\cal V}=V+\rt{4k_{{\rm B}}TR}\frac{\dwjp}{dt},
\eeq
so that
\beq
\dwj=\dwjp+dt\frac {\left(V-\an{V}\right)}{\sqrt{4k_{{\rm
B}}TR}}.
\label{dWdW'}
\eeq

Substitution of $P_{V}({\cal V})$ and $P({\cal V})$ into \erf{bayes}
gives to first order in $dt$
\beq
P_{{\cal V}}(V)
=  P(V)+\dwj\frac{V-\langle V\rangle}{\rt{4k_{{\rm B}}TR}}
P(V).
\label{hit}
\eeq
That is, we have derived the Kushner-Stratonovich equation
\beq
dP_{{\cal V}}(V)=
\dwj\frac{(V-\langle V\rangle)}{\rt{4k_{{\rm B}}TR}}P(V),
\label{PV}
\eeq
where the conditioning upon ${\cal V}$ is contained in
$\dwj=dt\left({\cal V}-\an{V}\right)/\rt{4k_{{\rm B}}TR}$.
This quantity is often referred to as the {\em residual} or {\em
innovation} \cite{KSBook,Dohe2000}.

\subsubsection{Joint Stochastic Equations}

To see how ${\cal V}$ conditions the quantum system, we form the
quantity $\rho(V)=\rho P(V)$,
where $\rho$ is here independent of $P(V)$ because we are still imagining
$I$ to be known at all times.
The time evolution of $\rho(V)$, given that ${\cal V}$ and $I$ are
known, is found from
\bqa
\rho(V)+d\rho_{I,{\cal V}}(V)&=&\left[P(V)+dP_{I}(V)+dP_{{\cal
V}}(V)\right] \nl{\times}
\left(\rho+d\rho_{I}\right),
\label{rhoCombo}
\eqa
with the use of \erf{PI},
\erf{PV} and \erf{rhoI}.

Substitution into \erf{rhoCombo} gives
\bqa
\rho_{I,{\cal V}}(V)&=&
\left[1+\dwj\frac{(V-\langle V\rangle)}{\rt{4k_{{\rm B}}TR}}
\right.
\nl{+}\left. dt\left(\frac{\partial}{\partial
V}\frac{V+IR} {RC}+\frac{{\rm P}\eta
e^{2}}{2\hbar\omega_{0}C^{2}}\frac{\partial^{2}} {\partial
V^{2}}\right)\right]P(V)
\nl{\times}\left[
1+dt\left({\cal L}+\sqrt{\eta}\xi(t){\cal
H}[e^{-i\Phi}c]\right)\right]\rho.
\label{long}
\eqa
The reader is reminded that $\xi(t)$ is uncorrelated with $\dwj$.
Upon using $I$ given in \erf{I} and noting that it contains the white
noise $\xi(t)$, the expansion of \erf{long} to $O(dt)$ is
\begin{widetext}
\bqa
\rho_{I,{\cal V}}(V)&=&
\left\{1+\dwj\frac{(V-\langle V\rangle)}{\rt{4k_{{\rm
B}}TR}}
+dt\left({\cal L}+\frac{\partial}{\partial
V}\frac{V} {RC}\right)\right.\nn\\
&&\left.+dt\frac{e}{C}\sqrt{\frac{{\rm P}}{\hbar\omega_{0}}}
\frac{\partial}{\partial
V}\left[\eta\langle e^{-i\Phi}c
+e^{i\Phi}c^{\dag}\rangle
+\rt{\eta}\xi(t)\right]\right\}\rho(V)\nn\\
 &&+ \left\{dt\rt{\eta}\xi(t)+[dt\xi(t)]^{2}
  \frac{e\eta}{C}\rt{\frac{{\rm P}}{\hbar\omega_{0}}}\frac{\partial}{\partial
  V}\right\}\nn\\
  &&\times\left[ e^{-i\Phi}c\rho(V)+
  e^{i\Phi}\rho(V)c^{\dag}-\an{e^{-i\Phi}c+
  e^{i\Phi}c^{\dag}}\rho(V)\right],
\label{longexp}
\eqa
\end{widetext}
where the form of the superoperator ${\cal H}$ in \erf{calH} has been
used.

\subsubsection{Average Over Unobserved Processes}

Finally, in reality ${\cal V}$ is known but $I$ is
not. Therefore we should average over the noise $\xi(t)$,
but keep the voltage noise $\dwj$.  The relevant averages are
\erf{xiSquare}, implying $dt^{2}{\rm E}[\xi(t)\xi(t)]=dt$.  An
important
consequence (necessary for consistency) is
a cancellation of the terms involving $\an{e^{-i\Phi}c+e^{i\Phi}c^{\dag}}$.
For convenience, we define a dimensionless voltage
$v=V\sqrt{C/4k_{{\rm B}}T}$, a rate $\gamma=1/RC$ and a dimensionless noise
power
\beq
N = \frac{4k_{{\rm B}}T\hbar\omega_{0}}{\eta R{\rm P}e^{2}}.
\label{NN}
\eeq
This last expression is the ratio of the Johnson noise power
($4k_{{\rm B}}TR$) to the low-frequency
power in ${\cal V}$ due the noise in the photocurrent
($R^{2}e^{2}{\rm P}\eta/\hbar\omega_{0}$).  The latter noise is that
which
would be present if a vacuum field was being combined with the LO
field at the homodyne detection beam splitter.

Averaging over $\xi(t)$ and using $v$, $\gamma$ and $N$  we then obtain the
following superoperator Kushner-Stratonovich equation
(SKSE) for
$\rho(v)$:
\bqa
d\rho_{{\cal V}}(v)&=&dt\left({\cal
L}+\frac{\gamma}{2N}\frac{\partial^{2}}{\partial
v^{2}}+\gamma\frac{\partial}{\partial v}v \right)\rho(v)
\nl{+}dt\frac{\partial}{\partial
v}\sqrt{\frac{\gamma\eta}{N}}
\left[e^{-i\Phi}c\rho(v)+
e^{i\Phi}\rho(v)c^{\dag}\right]
\nn\\&&+
\rt{\gamma}\dwj
\left(v-\langle v\rangle\right)\rho(v).
\label{dpHom}
\eqa
As noted below \erf{PV}, the conditioning on the quantity that is
measured in the laboratory, ${\cal V}$, is contained in $d{\cal W}_{{\rm
J}}(t)$, with
\beq
\rt{\gamma}\dwj=dt\gamma \left(\rt{\frac{C}{4k_{{\rm B}}T}}{\cal
V}-\an{v}\right).
\label{Vcondit}
\eeq

Although the description of the state of the photoreceiver (by $v$)
is essential in obtaining realistic trajectories, it is the evolution
of the quantum system that we are most interested in.  This state is
given by
\beq
\rho_{\cal V}=\int\rho(v)dv.
\eeq
 The average $\an{v}$, which
appears in \erf{dpHom}, is found from
\beq
\langle
v\rangle=\int dv\,{\rm Tr}[\rho(v)]v.
\eeq

\subsection{Effective Bandwidth}
\label{effBWTh}

As mentioned earlier, if there were no excess (Johnson)
noise added to the input signal, then this input signal
could be perfectly reconstructed from the filtered output.
However, in the case of
realistic detection presented above there is noise, characterized by
$N$, linearly added to the output of the filter.  The extent to which
information is lost due to this noise depends on the magnitude of the
noise, the filter bandwidth, and the nature of the evolution of the
monitored system.  In this subsection, we argue that in the limit of
small noise $N \ll 1$, the quality of
the photoreceiver can  be characterized by an
{\em effective bandwidth} that depends upon the noise and the filter
bandwidth. We expect the information loss to be small only if this
effective bandwidth is large compared to the relevant system
frequencies.

We identify the effective bandwidth, $B$, as being
roughly the frequency at which the noise in a vacuum input signal
is `lost' in the Johnson noise.  That is, the frequency at which the
power of the vacuum input signal becomes equal to the Johnson noise
power (which has a flat spectrum).    Obviously the loss of very high
frequency
noise will invariably occur in any practical photoreceiver.  The
question of whether this loss is important or not is answered by
looking at the eigenvalues of the Liouvillian.  If the input noise is
lost at frequencies that are well above the system rates then this
noise would not significantly affect the evolution of the system.
However, if noise is lost at frequencies at which the system can
respond to, then the purity of the system state will decrease considerably.
This is why we expect the effective bandwidth, as defined above, to
be the relevant parameter for the photoreceiver.

\begin{figure}
\includegraphics[width=\col]{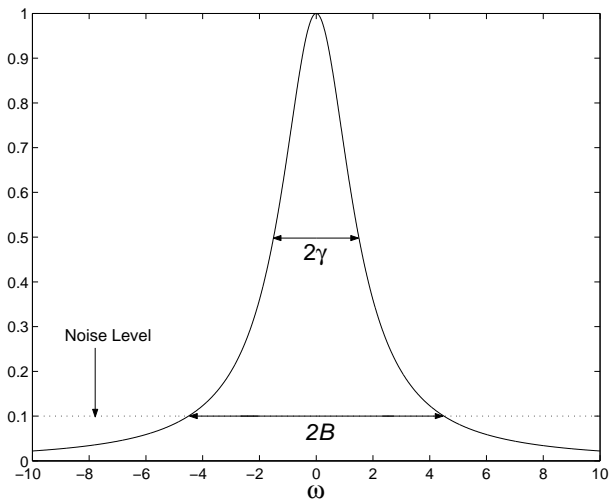}
\vspace{0.2cm}
\caption{The dimensionless quantities
$\gamma^{2}/(\gamma^{2}+\omega^{2})$ and $N$ are plotted, showing
their frequency dependence (or lack of it).  In this plot $\gamma=1.5$
and $N=0.1$.  The frequency at which the vacuum signal drops in power
to that of the noise is equal to the effective bandwidth $B$.}
\protect\label{EffectiveBWPlot}
\end{figure}

Equating the vacuum signal and Johnson noise powers at the frequency
$\omega = B$ gives
\bqa
\frac{\gamma^{2}}{\gamma^{2}+B^{2}} = \frac{4k_{{\rm B}}TR\hbar\omega_{0}}
{R^{2}e^{2}{\rm P}\eta} = N
\label{gamN}
\eqa
This is shown graphically in Fig.~\ref{EffectiveBWPlot}.  Obviously
if $N$ is too large then this equation has no solution in the real numbers,
 and indeed
the intuitive picture behind the argument fails. But for small $N$ we
have
\beq
B = \frac{\gamma\rt{1-N}}{\rt N}\approx\frac{\gamma}{\rt N}.
\label{effBW}
\eeq
  This result will be
investigated further for two different quantum systems in the following paper.

In terms of physical parameters
\beq
B=\frac{e}{2C}\rt{\frac{\eta {\rm P}}{k_{{\rm
B}}TR\hbar\omega_{0}}}.
\eeq
The photon flux ${\rm P}/\hbar\omega_{0}$ of the LO cannot be increased
arbitrarily as the regime in
which the photodiode responds linearly to the field incident upon it must be
adhered to.  The resistor $R$ cannot be reduced to zero as it is necessary
for the reduction of the effective input impedance of the op-amp so that the
photodiode can act as a current source.  It also converts the current to a
voltage, which presumably would be swamped by other, not considered, noise
sources if reduced too far.  Minimizing stray capacitances and the
temperature of operation  will
improve the photoreceiver operation.

\subsection{Adding Noise Only}
\label{NoiseOnly}
The limit in which the power of the Johnson noise goes to zero has
already been discussed, with the result being perfect detection.
If the response time of the RC circuit
vanishes then, provided Johnson noise is still added, the input signal
is still obscured.  For this to be a physically sensible consideration
 we assume that
the capacitance, rather than the resistance, is zero.

The output voltage of the photoreceiver is then given by
\beq
{\cal V}= -IR+\rt{4k_{{\rm B}}TR}\frac{\dwjp }{dt},
\eeq
as indicated by the form of \erf{Vo}.  Here $\dwjp/dt$ represents the
`real' Johnson noise which is of a Gaussian nature.  Using \erf{I}
and \erf{NN} gives (with $\Phi=0$ for convenience),
\bqa
{\cal V}&\propto&{\frac{\eta}{1+N}}\langle c+c^{\dag}\rangle+
\sqrt{\frac{\eta}{1+N}}\theta(t),
\label{four}
\eqa
where $\theta(t)$ is normalized white noise consisting of a linear
combination of the vacuum
input noise  and the Johnson noise.
 Comparison with \erf{I} shows that the effect of adding noise but no filtering
 is to reduce the efficiency by a factor
of $1+N$. It can be shown using Bayesian analysis that this result
holds for the form of the quantum trajectory as well.

\section{Conclusion}
\label{Concl}

The evolution of open quantum systems conditional upon detection 
results from {\em realistic} detectors cannot in general be generated 
by standard quantum trajectory theory (stochastic master equations). 
A method for generating this evolution was proposed in 
Ref.~\cite{WarWisMab01}. The crucial element is to include 
correlations between the system and classical 
detector states which cannot be observed in practice. 
In this paper we have given a full 
description of this method, and shown how it can be applied in 
quantum optics. In particular, we derive realistic quantum 
trajectories for conditioning upon photon counting using an 
avalanche photodiode, and homodyne detection using a photoreceiver.
These equations were presented in Ref.~\cite{WarWisMab01}, but a full 
derivation was not given there.

In this paper we have not provided any solutions of 
the equations we have derived. To find and study these solutions is 
not a minor task, which is why it is reserved for the
 following paper \cite{WarWis02b}. There we use our equations  
to determine the evolution of a two-level cavity QED 
system, conditioned on four different types of detection (using the 
two detectors mentioned above). We also consider another system which 
can be treated analytically for realistic homodyne detection. 
Our study achieves five important aims, 
which are outlined in the introduction to that paper.

It is worth re-emphasizing the generality of our approach. It is 
applicable not just in quantum optics, but in all areas where open 
quantum system theory is used. Prime examples are the mesoscopic 
measurement devices used in quantum electronics, such as single 
electron transistors. This device has received much attention as a
possible measurement device for solid state qubits
\cite{WisWah01,JJQubit,Dev00}. Conditional states may be used in quantum 
computation both for preparation, and for non-deterministic gate 
implementation \cite{NieChu00}. 
In this context, the necessity of being able to relate
realistically available results to the state of the qubit is obvious.

The greatest significance of our work is in the field of quantum 
control. The conditional state of a quantum system is synonymous with 
an observer's knowledge about that system. Thus (assuming all 
uncertainties are properly taken into account)  it is by definition 
the optimum mathematical object to use to control that system. 
Properly taking into account detector imperfections is essential to 
building optimal control loops. We thus expect our theory to have 
broad applications in future quantum technology.

\end{document}